# Broken symmetries associated with a Kagome chiral charge order

**Authors:** Zi-Jia Cheng[1]*, Md Shafayat Hossain[1]*†, Qi Zhang[1]*, Sen Shao[2]*, Jinjin Liu[3,4,5]*, Yilin Zhao[2], Mohammad Yahyavi[2], Yu-Xiao Jiang[1], Jia-Xin Yin[1], Xian Yang[1], Yongkai Li[3,4,5], Tyler A. Cochran[1], Maksim Litskevich[1], Byunghoon Kim[1], Junyi Zhang[6], Yugui Yao[3,4], Luis Balicas[7], Zhiwei Wang[3,4,5]†, Guoqing Chang[2]†, M. Zahid Hasan[1]†

**Affiliations:**
[1]Laboratory for Topological Quantum Matter and Advanced Spectroscopy, Department of Physics, Princeton University, Princeton, New Jersey, USA.
[2]Division of Physics and Applied Physics, School of Physical and Mathematical Sciences, Nanyang Technological University, 21 Nanyang Link, 637371, Singapore.
[3]Centre for Quantum Physics, Key Laboratory of Advanced Optoelectronic, Quantum Architecture and Measurement (MOE), School of Physics, Beijing Institute of Technology, Beijing 100081, China.
[4]Beijing Key Lab of Nanophotonics and Ultrafine Optoelectronic Systems, Beijing Institute of Technology, Beijing 100081, China.
[5]Material Science Center, Yangtze Delta Region Academy of Beijing Institute of Technology, Jiaxing 314011, China.
[6]Institute for Quantum Matter and Department of Physics and Astronomy, Johns Hopkins University, Baltimore, Maryland 21218, USA.
[7]National High Magnetic Field Laboratory, Tallahassee, Florida 32310, USA.

†Corresponding authors, E-mail: mdsh@princeton.edu; zhiweiwang@bit.edu.cn; guoqing.chang@ntu.edu.sg; mzhasan@princeton.edu.

*These authors contributed equally to this work.

**Abstract:**
**Chirality or handedness manifests in all fields of science, ranging from cell biology, molecular interaction, and catalysis to different branches of physics. In condensed matter physics, chirality is intrinsic to enigmatic quantum phases, such as chiral charge density waves and chiral superconductivity. Here, the underlying chiral response is subtle and leads to broken symmetries in the ground state. Detection of subtle broken symmetries is the key to understand these quantum states but they are extremely challenging to expose leading to debate and controversy. Here, using second-order optical response, we uncover the broken symmetries of a chiral charge density wave in the Kagome lattice $KV_3Sb_5$, revealing the relevant broken symmetries of its charge order. $KV_3Sb_5$ undergoes a phase transition to a charge-ordered state at low temperatures. Our polarization-dependent mid-infrared photocurrent microscopy reveals an intrinsic, longitudinal helicity-dependent photocurrent associated with the charge order. Our measurements, supported by our theoretical analysis, provide direct evidence for broken inversion and mirror symmetries at the charge order transition, indicating a chiral charge ordered state. On the other hand, we do not observe a circular photogalvanic effect along the direction perpendicular to that of the incident light, imposing stringent constraints on the rotational and point group symmetries of the charge order. Our study not only visualizes the chiral nature of the Kagome charge order revealing its broken**

**symmetries, but also highlights the nonlinear photogalvanic effect as a sensitive probe for detecting subtle symmetry breakings.**

**Main Text:**

In the framework of Ginzburg–Landau theory, the phase transitions result from symmetry breaking of the ground state and associated order parameters[1]. Therefore, scrutinizing the broken symmetries in the post-transition state is imperative for comprehending the mechanism driving the phase transition, particularly in systems where multiple interaction pathways are involved. For instance, charge density wave (CDW) like instabilities, where the electron density spontaneously undergoes periodic modulations, have been found to be widely present in high-temperature superconductors and strongly intertwined with the superconductivity[2–4]. The microscopic structure of the intertwined charge orders in strongly correlated systems, including the symmetry and anisotropy, provides invaluable insight in the interactions between charge, lattice and spin degrees of freedom and therefore has been attracting significant research attention[5–7].

Recently, new Kagome materials $AV_3Sb_5$ (A = K, Rb, Cs) have become a major focus of research endeavors, in which intertwined CDW and superconducting orders emerge at low temperatures[8–10]. Below the CDW transition temperature ($T_c$ = 79 K – 102 K), the vanadium atoms form a star-of-David pattern in the Kagome plane and the in-plane wave (Q) vectors connecting the van-Hove singularities at different M points, induce a strong electronic band reconstruction[11,12]. Intriguingly, scanning tunneling microscopy (STM)[13], muon spin relaxation[14], optical Kerr[15], and electrical transport measurements[16] provided evidences for the spontaneous time-reversal symmetry breaking of the charge ordered phase and the long-thought loop current state, although contradicting results were also reported[17–19]. In parallel, scanning birefringence microscopy[15] and STM[17] studies revealed the broken six-fold rotation symmetry and the presence of nematicity in the charge order state, which are challenged by the recent thermodynamic measurements[20]. The CDW was further demonstrated to be coexisting with superconductivity at lower temperatures[21] and forms an electronic smectic state. However, despite intensive research, the exact symmetry of the highly unconventional charge-ordered states and the associated crystal structure is still under strong debate and remains elusive, partially due to the weak lattice distortion and the abundance of energy degenerate CDW configurations. Here, using the nonlinear photogalvanic effect, we fully elucidate the symmetries of the enigmatic charge order in $KV_3Sb_5$.

The circular photogalvanic effect (CPGE), which measures the change of second-order photocurrent response under light with opposite circular polarizations, is a powerful tool for detecting broken symmetries[22,23] and Berry curvature distributions [24–26]. The corresponding response tensor inherits the symmetry of the system and the CPGE with long-wavelength probing light (for example mid(far) infrared) can directly reveal the microscopic symmetry breaking of the wavefunctions of the low-energy states[22,27]. Remarkably, in $KV_3Sb_5$ we observe the emergence of CPGE upon the onset of the CDW order. The intrinsic CPGE directly exposes the broken mirror and inversion symmetries within the charge order. The absence of CPGE in the horizontal direction further imposes constraints on the rotational and point group symmetries of the charge order at low temperatures. Thus, our study reveals the symmetries of the unconventional CDW order in $KV_3Sb_5$ and lays the foundations for further understanding of the order-parameter in charge-ordered phases.

We employed a mid-infra-red (10.6 μm wavelength, ~120 meV photon energy) scanning photocurrent microscope to investigate the polarization-dependent photogalvanic response in KV$_3$Sb$_5$. The experimental setup is schematically illustrated in Fig. 1**a**, where the polarization of the mid-infra-red laser beam is modulated using a quarter-wave plate prior to being directed perpendicularly to the sample surface. To measure the longitudinal (sample's out-of-plane) photocurrent, we fabricated a sandwich structure consisting of graphite (as an electrode), KV$_3$Sb$_5$, and metal electrodes, as shown in Fig. 1**b**. The current generated across the top graphite electrode and the bottom metallic electrode was measured. The optical microscopy image of the device, including multiple electrodes, is presented in Fig. 1**c**.

We begin our investigation by measuring spatially resolved photocurrent at $T = 10$ K to identify the regions where the photocurrent is generated. The resulting map, depicted in Fig. 1**d**, reveals a prominent positive photocurrent source at the intersection region between graphite and KV$_3$Sb$_5$. Additionally, a small negative photocurrent appears at the interface between exposed KV$_3$Sb$_5$ and the bottom electrode. This uneven and bipolar distribution of photocurrent suggests the presence of out-of-plane photo-thermoelectric currents stemming from the variation in the Seebeck coefficient at the contacts[25,28]. Such a distribution allows us to differentiate the contribution of CPGE from those associated with photo-thermoelectric currents in real space, as we will discuss further. Next, we focused the beam at the center of the positive photocurrent area (marked by the green star in Fig. 1**d**) and recorded the out-of-plane photocurrent ($I_z$) as a function of the azimuthal angle of the quarter-wave plate. Remarkably, we find that the photocurrent generated under left-circular-polarized pump light is greater than that under right-circular-polarized light. This difference is the hallmark of the CPGE. To dissect the different components contributing to the photocurrent, we employed a fitting procedure using the following function[29]:

$$I_z = A \sin 2\phi + B \sin 4\phi + C \cos 4\phi + D, \tag{1}$$

where $\phi$ represents the quarter-wave plate angle. The fitting results presented in Fig. 1**f** reveals the presence of both the CPGE (A term, exhibiting a periodicity of 180 degrees) and linear photogalvanic effect (LPGE) contributions (B and C terms, exhibiting a periodicity of 90 degrees). Additionally, a polarization-independent contribution from photo-thermoelectric currents is observed.

To elucidate the origins of the different photocurrent contributions, we investigated their temperature dependence. Figure 2**a** presents a series of polarization-dependent measurements acquired at the same location but at different temperatures. At low temperatures, and as discussed earlier, a distinct CPGE signal emerges, allowing us to define $I_{CPGE}$ as half of the signal difference between left-circular-polarized and right-circular-polarized light. As the temperature increases, $I_{CPGE}$ gradually becomes less prominent before sharply diminishing above $T = 90$ K. To quantitatively analyze the temperature dependence, we extracted the value of this nontrivial component at each temperature by fitting the data using Eq. 1. The results are summarized in Fig. 2**b** for the CPGE component and Fig. 2**c** for the LPGE component. We find that $I_{CPGE}$ follows a power-law behavior below a critical temperature ($T^* \simeq 95 \pm 10$ K, close to the CDW transition temperature) and saturates at low temperatures with a magnitude of 3 nA. On the other hand, the helicity-independent component ($\sqrt{B^2 + C^2}$) exhibits a gradual decrease with rising temperatures, likely due to the increasing resistivity of the system. This is consistent with the trend expected for the photo-thermoelectric effect and does not show any obvious anomaly near $T^{*}$[30]. The stark disparity in the temperature dependencies strongly supports a non-thermal origin for the CPGE (see also Extended Fig. 1). Instead, it suggests the presence of a new electronic order characterized by broken symmetries, which accounts for the observed CPGE effect at lower temperatures.

Having observed the CPGE associated with the charge order, we investigate its underlying mechanisms. To this end, we conducted a series of experiments to explore its correlation with the laser power, cooling conditions, contact geometry, and spatial distribution. The linear power dependence of the CPGE, as demonstrated in Extended Fig. 6, is consistent with its characterization as a second-order effect; see additional discussion in Methods Section VIII. We further cooled the device from $T = 120$ K to 20 K using either right-circular-polarized or left-circular-polarized light with an illumination power of 2.9 mW. Subsequently, we measured the polarization dependence of $I_z$ at the same location. The results are depicted in Fig. 3**a**, where the two curves exhibit close quantitative alignment. Importantly, $I_z$ shows no apparent dependence on the "induction process," distinguishing itself from the previously observed chiral induction and gyrotropic charge density wave (CDW) phase in 1T-TiSe$_2$[22]. Furthermore, we investigated the in-plane photocurrent at $T = 10$ K and found no detectable polarization dependence (Fig. 3**b**). This observation suggests that the CPGE predominantly manifests along the longitudinal direction when subjected to the normal incidence of light. Such direction sensitivity rules out extrinsic influences such as fabrication-related anomalies or variations in input power with the azimuthal angle of the quarter-wave plate rotation as potential sources of the observed CPGE. Lastly, we examined the spatial variability of the CPGE by analyzing the half-difference between the $I_z$ measured under right-circular-polarized and left-circular-polarized light as we scanned the beam across the device (Figs. 3**c** and 3**d**). Interestingly, the CPGE signal measured at $T = 10$ K is primarily localized in the region where the top graphite overlaps KV$_3$Sb$_5$, in contrast to the spatial mapping of $I_z$. At $T = 120$ K, the CPGE signal becomes undetectable across the entire sample. This distinct spatial distribution of the CPGE current further discredits the thermal effect as the underlying generation mechanism[31]. Collectively, these experiments shed light on the CPGE's behavior and rule out potential extrinsic factors, providing compelling evidence for an intrinsic origin of the observed CPGE in the graphite-KV$_3$Sb$_5$ device.

Having discussed our experimental observations, we turn to explore the plausible mechanisms and symmetry prerequisites underlying the intriguing longitudinal CPGE in KV$_3$Sb$_5$. As the spin-orbit coupling in KV$_3$Sb$_5$ is weak[8] and long-range magnetic order is absent[32], the spin photogalvanic effect is expected to be minimal in KV$_3$Sb$_5$. Therefore, the observed helicity-dependent photocurrent is mainly driven by the orbital excitations, which in turn inherit the global symmetry of the electronic structure.

The photocurrent contributed from a ***k***-point in the momentum space is determined by the CPGE tensor $\tilde{\beta}_{ij}(\boldsymbol{k})$, where $i$ represents the current direction and $j$ denotes the laser propagating direction. $\tilde{\beta}_{ij}(\boldsymbol{k})$ is proportional to the product of the Fermi velocity $v_i(\boldsymbol{k})$ and the Berry curvature $\Omega_j(\boldsymbol{k})$ [33]. Initially, we focus on the impact of inversion symmetry on the CPGE. For a generic momentum point ***k*** and its partner −***k*** linked by inversion symmetry, the Fermi velocities will change the signs $v_i(-\boldsymbol{k}) = -v_i(\boldsymbol{k})$, while the Berry curvatures remain unchanged $\Omega_j(-\boldsymbol{k}) = \Omega_j(\boldsymbol{k})$. Consequently, the resulting photocurrents from each pair of inversion partner cancel each other out, with $\tilde{\beta}_{ij}(-\boldsymbol{k}) = -\tilde{\beta}_{ij}(\boldsymbol{k})$ indicating that breaking inversion symmetry is essential for KV$_3$Sb$_5$ to exhibit a nonzero CPGE.

Next, we examine the influence of mirror planes $M$ on the CPGE within KV$_3$Sb$_5$, starting with $M_x$, which transforms $\boldsymbol{k} = (k_x, k_y, k_z)$ into $M_x\boldsymbol{k} = (-k_x, k_y, k_z)$. This operation only alters the sign of $k_x$, leading to a sign change in the x-component of the velocity while leaving the others unaffected: $v_x(M_x\boldsymbol{k}) = -v_x(\boldsymbol{k})$, $v_y(M_x\boldsymbol{k}) = v_y(\boldsymbol{k})$, $v_z(M_x\boldsymbol{k}) = v_z(\boldsymbol{k})$ (Fig. 4**b**, left-top). On the other hand, Berry curvature, being a pseudovector, only maintains its components perpendicular to the mirror plane, resulting in $\Omega_x(M_x\boldsymbol{k}) = \Omega_x(\boldsymbol{k})$, $\Omega_y(M_x\boldsymbol{k}) = -\Omega_y(\boldsymbol{k})$, $\Omega_z(M_x\boldsymbol{k}) = -\Omega_z(\boldsymbol{k})$ (Fig. 4**b**, left-bottom). In scenarios where the laser propagates along the z-axis and the current is measured in the same direction, the presence of a mirror plane $M_x$ results in a null net current, as $\tilde{\beta}_{zz}(M_x\boldsymbol{k}) = v_z(\boldsymbol{k}) \times [-\Omega_z(\boldsymbol{k})] = -\tilde{\beta}_{zz}(\boldsymbol{k})$. A similar rationale concludes that mirror plane $M_y$ (or $M_z$) also eliminates $\tilde{\beta}_{zz}$, as illustrated in Fig. 4**b** (right). Therefore, to observe the nonzero photocurrents reported in our study $\tilde{\beta}_{zz}$, all mirror planes must be broken in the correlated phases of KV$_3$Sb$_5$.

Taking all the above into consideration, we experimentally demonstrated that all the inversion and mirror symmetries are broken in KV$_3$Sb$_5$ at low temperatures. Subsequently, we turn to determining the potential point groups of chiral KV$_3$Sb$_5$. Our measurements, as illustrated in Fig. 3, revealed a polarization-dependent in-plane CPGE of zero, indicating that, $\tilde{\beta}_{xz}$ and $\tilde{\beta}_{yz}$ are both zero. This outcome stems from the rotational symmetries along the z-axis present in KV$_3$Sb$_5$. Specifically, any of the following rotational symmetries along the z-axis would nullify the in-plane CPGE: $C_{2z}$, $S_{2z}$, $C_{3z}$, $S_{3z}$, $C_{6z}$, and $S_{6z}$ (see Methods and Extended Fig. 3). Therefore, based on our measurements, we can infer that the possible point groups for the CDW phase of KV$_3$Sb$_5$ are $C_2$, $D_2$; $C_3$, $D_3$; $C_6$; $D_6$. Considering the evidence of a two-fold rotation charge distribution on the surface in previous studies[13,17], $C_2$ and $D_2$ are the most possible point groups. However, we cannot completely rule out the possibility that the bulk of chiral KV$_3$Sb$_5$ still preserves higher-fold screw rotations $S_{3z}$ and or/and $S_{6z}$, which may not be detectable using surface-sensitive techniques.

We would like to further address further the challenge of pinning down the structural chirality in KV$_3$Sb$_5$. Previous STM measurements observed a rotating surface charge distribution at low temperatures, suggesting a chiral charge order in KV$_3$Sb$_5$[13]. However, this rotating surface charge distribution only confirms the breaking of $M_x$ and $M_y$. For instance, by applying the $M_z$ operation, we can map a left-handed crystal into a right-handed crystal while the in-plane rotation direction remains the same (Fig. 4**c**). After the combination of the two opposite chiral crystals, the in-plane rotation remains unchanged, but the crystal is now an achiral structure with $M_z$. In other words, surface-sensitive experiments cannot fully rule out the existence of $M_z$. In another recent electronic magneto-chiral anisotropy measurement, the author also concluded that they can only confirm the breaking of $M_x$ and $M_y$ but not $M_z$[16]. In the current experiment, the long penetration depth of the 10 μm incident light (~ 20 nm) and the thickness of the device (~100 nm) ensures that the symmetry of the bulk crystal is probed, while the asymmetry induced by the interfaces is likely confined to a few layers at the interface and therefore is expected to be dominated by the bulk response. We note that the small atom displacement in KV$_3$Sb$_5$ is the main reason challenging the experimental determination of the structure of KV$_3$Sb$_5$. Here, we compare KV$_3$Sb$_5$ with another CDW material TaTe$_2$[34]. The average displacement is around 0.15 Å in TaTe$_2$, while the displacement in KV$_3$Sb$_5$ is only around 0.033 Å. This difference can be visualized in our simulated XRD patterns. Detectable alterations can be observed between the high-temperature and low-temperature structures of TaTe$_2$ (Fig. 4**d**). In contrast, the XRD patterns of the two phases of KV$_3$Sb$_5$ are nearly identical (Fig. 4**e**). Our systematic photocurrent study thus provides strong constraints on the symmetries of the chiral charge order in KV$_3$Sb$_5$.

Our experimental observations of longitudinal CPGE in $KV_3Sb_5$ associated with its charge order provide direct evidence for broken inversion and mirror symmetries, indicating that the bulk CDW order is chiral, namely the electronic state is not superimposable on its mirror image. Although the atomic displacements in the charge-ordered state are small (Fig. **4d**), the electron-phonon coupling may contribute to the symmetry breaking observed in $KV_3Sb_5$. Our phonon calculations for the pristine phase of $KV_3Sb_5$ reveal softened phonon modes along the path from M to L (Extended Fig. 7). This indicates that electron-phonon interactions could be a driving force behind the CDW phase transition in $KV_3Sb_5$, aligning with the findings from previous studies[35,36]. Conversely, the detection of saddle points and quasi-nested electron and hole pockets on the Fermi surface points to inherent Fermi-surface instabilities[11]. The high tunability of this intriguing CDW phase further underscores the potential role of electron-electron interactions, suggesting that they are also critical in the formation of the chiral CDW instability[13,16,37,38]. Therefore, the CDW phase in $AV_3Sb_5$ (A = K, Rb, Cs) may result from a complex interplay between electron-phonon and electron-electron interactions. While this study provides a thorough investigation of the CPGE response in $KV_3Sb_5$, comparing it with other members of the metallic Kagome family, such as $CsV_3Sb_5$ and $RbV_3Sb5$, remains an important direction for future research. This is particularly relevant given the differences in chiral transport properties observed across the family[39]. Furthermore, our approach utilizing non-linear photocurrent response can serve as a powerful tool for detecting subtle intrinsic symmetry breakings that lead to chiral electronic orders not only in other Kagome compounds but also in broader range of quantum materials.

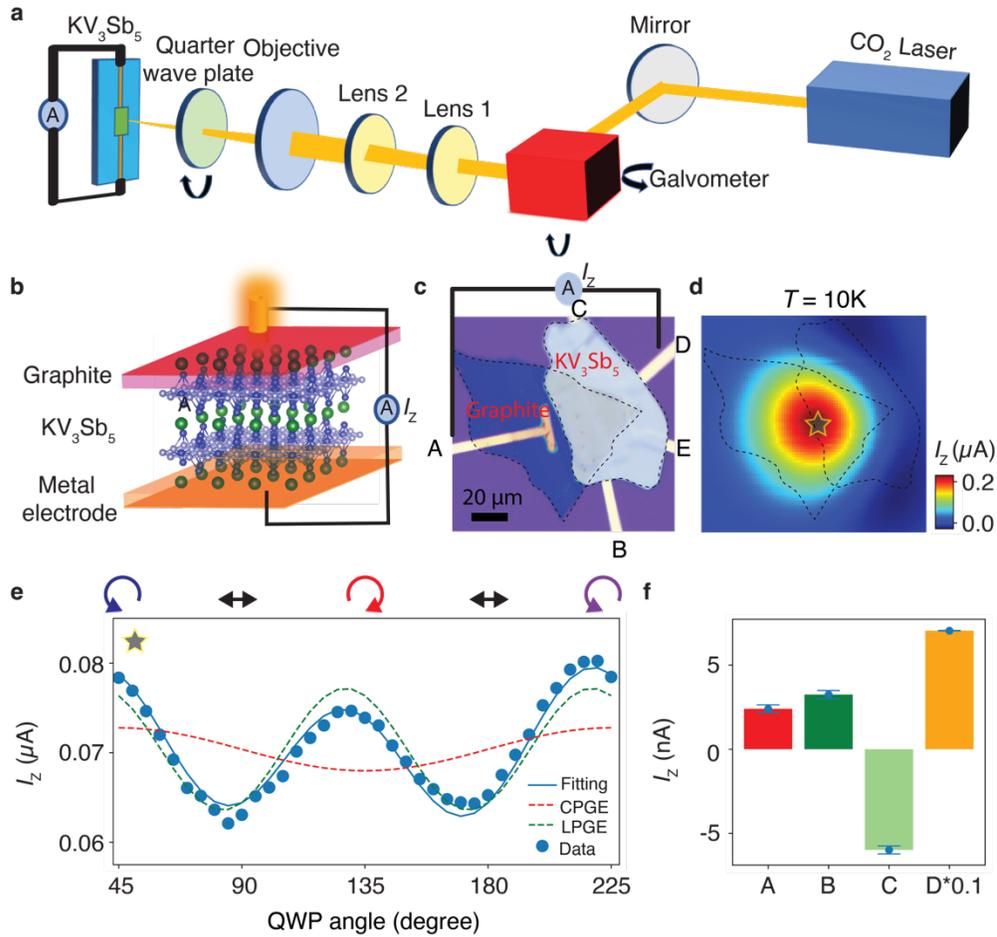

**Fig. 1. Observation of the longitudinal circular photogalvanic effect in $KV_3Sb_5$.** **a**, Overview of the mid-IR photocurrent microscope setup. **b**, Schematic illustrating the device configuration for measuring the out-of-plane photocurrent ($I_z$). **c**, Optical microscopy image of the device. The electrodes are labeled as A-E. **d**, Spatial mapping of the $I_z$ photocurrent, with the largest signal observed at the intersection between graphite and $KV_3Sb_5$. **e**, Polarization dependence of $I_z$ when the beam spot is positioned at the marked star in **d**. The pump light polarization evolves from left-circular (LC) to horizontal (LH) to right-circular (RC) to horizontal (LH) and back to left-circular (LC) as the QWP angle changes from 45 to 225 degrees. **f**, Decomposition of the photocurrent into four different channels.

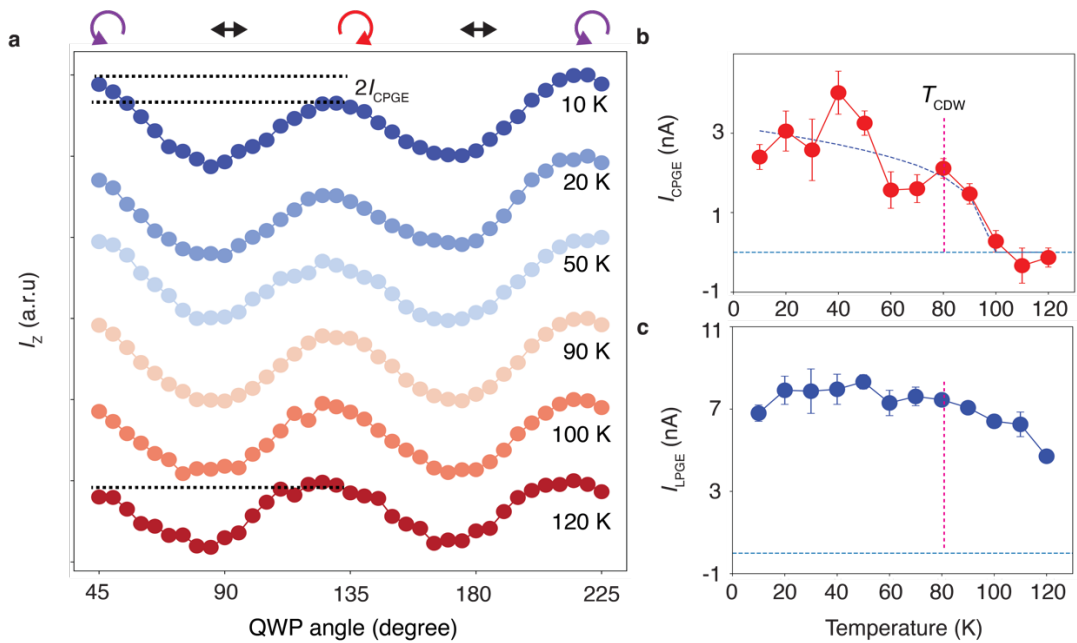

**Fig. 2. Temperature dependence of the longitudinal photocurrents highlighting a correlation with the CDW transition. a**, Polarization dependence of $I_z$ at different temperatures. The curves are normalized and offset for better comparison. **b**, Temperature dependence of the circular photogalvanic effect (CPGE). **c**, Temperature dependence of the linear photogalvanic effect (LPGE) component. It is worth noting that the CPGE nearly vanishes above $T = 90$ K, while the LPGE shows no distinguishable change across the charge density wave transition ($T_{CDW}$). $T_{CDW}$ is extracted from measurements of the resistivity as a function of the temperature (see Extended Fig. 2) performed in a bulk $KV_3Sb_5$ crystal.

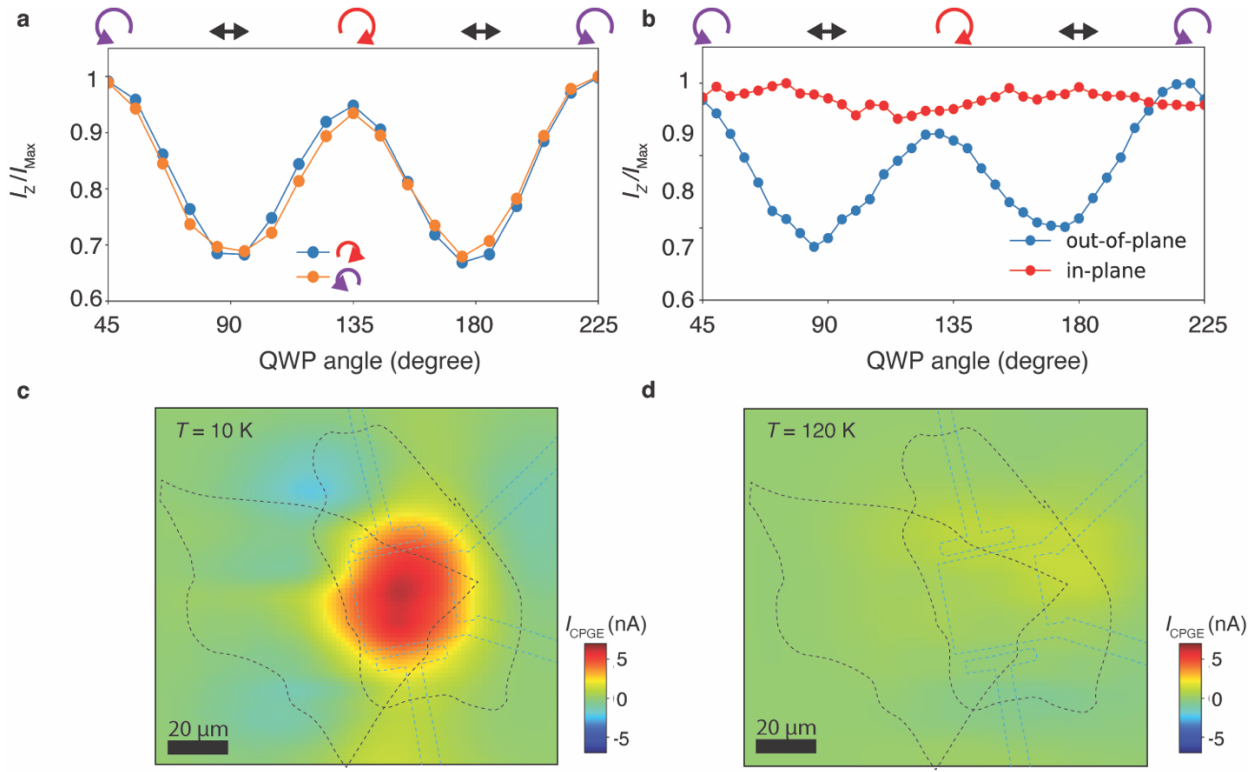

**Fig. 3. Investigation on the origin of the circular photogalvanic effect. a**, Polarization dependence of the normalized out-of-plane photocurrent measured on the same device which was cooled down with either right-circular (blue points) or left-circular polarized (orange points) light. **b**, Out-of-plane (blue) and in-plane (red) photocurrent as a function of light polarization. The measurement is performed at $T = 10$ K. For the in-plane photocurrent measurement, all contacts except contacts C and B (as labeled in Fig. 1**c**) are grounded. **c**, CPGE distribution measured as a function of the beam spot location at $T = 10$ K. The major contribution of CPGE appears circular, consistent with prior reports of CPGE from other materials[22,31], and occurs at the overlap region between graphite, $KV_3Sb_5$ and the bottom contact (blue dashed line). **d**, CPGE distribution measured as a function of the beam spot location at $T = 120$ K.

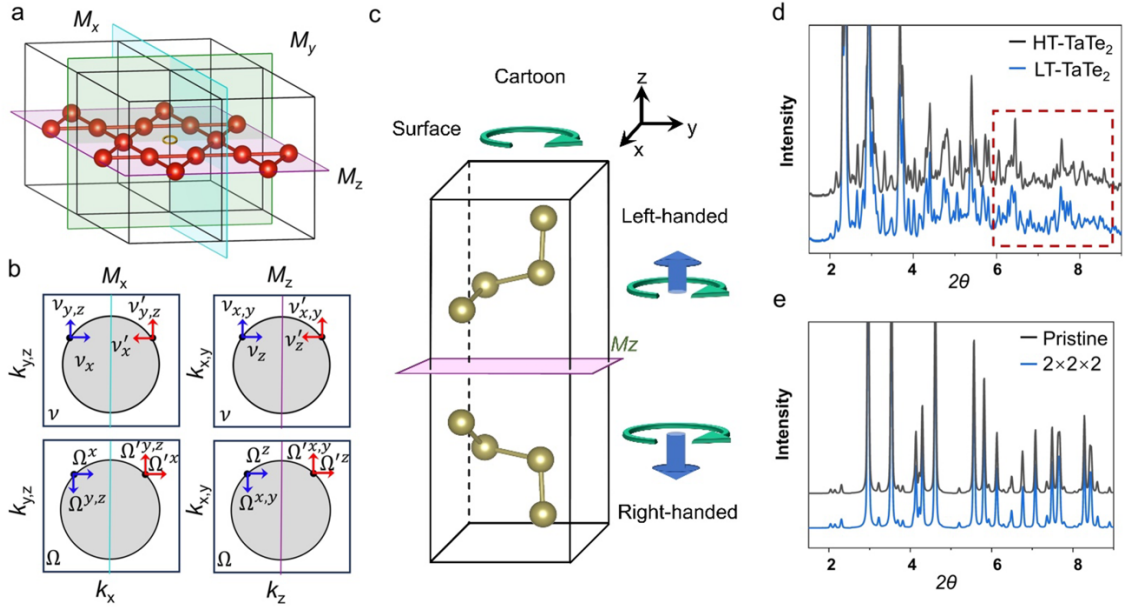

**Fig. 4. Symmetry breaking induced Chirality in KV$_3$Sb$_5$**, **a,** Pristine phase of KV$_3$Sb$_5$, preserving inversion and mirror symmetries. **b,** Illustration of the effects of mirror symmetries on velocity and Berry curvature. **c,** Top and bottom patterns depicting structures with opposite chirality. The combination of two different chiral structures in a crystal results in an achiral structure with $M_z$ mirror symmetry. **d,** Calculated angle-resolved X-ray diffraction of high temperature (HT) and low temperature (LT) phase for TaTe$_2$ **e,** Calculated angle-resolved X-ray diffraction of the pristine phase and CDW phases of KV$_3$Sb$_5$.

# Methods:

### I. Single crystal synthesis:

$KV_3Sb_5$ single crystals were grown by the flux method. K (99.95%), V (99.999%), and Sb (99.9999%) with the K/V/Sb=7: 3: 14 molar ratio were fully mixed. The mixture was then placed into an alumina crucible, sealed in a quartz tube with $10^{-3}$ Pa of Ar pressure, heated up to 1000°C in 200 h, kept at 1000°C for 12 h, and subsequently cooled to 200°C over 230 h. After the stove came down to room temperature, the flux was removed by putting the mixtures into deionized water. Finally, the high-quality single crystals of $KV_3Sb_5$ were harvested.

### II. Device fabrication:

We utilized a polydimethylsiloxane (PDMS) stamp-based mechanical exfoliation technique to fabricate $KV_3Sb_5$ devices. We patterned the electrical contacts onto the sample exfoliated onto the silicon substrates with a 280 nm layer of thermal oxide using electron beam lithography, followed by chemical development and metal deposition (5 nm Cr/35 nm Au). The fresh $KV_3Sb_5$ crystals were mechanically exfoliated from bulk single crystals on PDMS stamps. Prior to transferring them onto the $SiO_2$/Si substrates with pre-patterned Cr/Au electrodes, we visually inspected the crystals under optical microscopy to select samples with favorable geometries. Note that the $KV_3Sb_5$ flake was transferred on the top of the pre-patterned contacts labeled B-E in Fig. 1**c**. The in-plane photocurrent (vertical to the light incidence) of the $KV_3Sb_5$ device was measured between electrodes B-E. To measure the photocurrent perpendicular to the sample plane (along the incident light), a few-layer graphene (graphite) was introduced as a top contact to connect the electrode A and the $KV_3Sb_5$ flake. The entire sample fabrication process was conducted within a glovebox equipped with a gas purification system, maintaining an environment with extremely low levels of oxygen and water vapor (<1 ppm).

### III. Photocurrent measurements:
The probing light for our experiments was generated from a $CO_2$ laser (Access Laser) and modulated using a mechanical chopper (Thorlabs) operating at a frequency of 167 Hz. To focus the laser onto the sample, we employed a ZnSe lens with a focal length of 50 mm, resulting in a beamspot size of approximately 25 μm at full width at half maximum. The input power on the sample surface was maintained at 2.9 mW. To detect the current signal, we employed a current preamplifier (SR570) for initial amplification, followed by the measurements using a lock-in amplifier (SR830). This setup ensured accurate and sensitive current detection. The device under investigation was cooled down using a customized cryostat (Oxford MicroHires) specifically designed for our experimental requirements. To ensure the reliability and consistency of our measurements, we repeated the experiments on multiple devices and performed multiple measurements for each device to ensure the reproducibility of the results.

### IV. First-principles calculations based on X-ray diffraction
The structures of the pristine phase and 2×2×2 CDW phase of $KV_3Sb_5$ are based on previous experimental work[8] and theoretical predictions[40], respectively. To minimize structural disparities between experimental data and theoretical calculations, we performed structural relaxations using identical parameters. Our *ab-initio* calculations were performed within the framework of density functional theory (DFT), employing the Vienna Ab initio Simulation Package (VASP) with the Perdew−Burke−Ernzerhof exchange−correlation functional [41–43]. The valence electrons for K, V, and Sb were treated as $4s^1$, $3d^34s^2$, and $5s^25p^3$, respectively. In our calculations, we used a kinetic energy cutoff of 520 eV and a $k$-spacing parameter of 0.2 to ensure energy convergence within 1 meV/atom. An energy threshold of $10^{-6}$ eV and a force threshold of $3\times10^{-3}$ eV/Å were applied. To account for van der Waals interactions, we included the zero-damping DFT-D3 method[44] in our calculations. The simulated angle-resolved X-ray diffraction (XRD) data for the pristine phase and 2×2×2 CDW phase in the $KV_3Sb_5$ system were generated with an X-ray wavelength of 0.1671 Å[9]. For the $TaTe_2$ system, both the high-temperature and low-temperature phase structures were experimentally determined and directly used for simulating X-ray diffraction data, with an X-ray wavelength of 0.1173 Å[45].

### V. Symmetry analysis of in-plane circular photogalvanic effect tensors

Without loss of generality, we consider the expression of the CPGE tensor, $\tilde{\beta}_{ij}(\mathbf{k})$, for a two-band model, which is proportional to the product of the Fermi velocity $v_i(\mathbf{k})$ and the Berry curvature $\Omega_j(\mathbf{k})$ .[33]

**Symmetry analysis of in-plane circular photogalvanic effect tensors**

- $C_{2z}$ rotational symmetry:

Under the symmetry operation $C_{2z}$, a generic point $k$ is related to $k' = C_{2z}k = (-k_x, -k_y, k_z)$. The Fermi velocities and Berry curvature obey the following constraints (Extended Fig. 3 (a-b)).

$$\begin{cases} v_x(C_{2z}k) = -v_x(k) \\ v_y(C_{2z}k) = -v_y(k) \\ v_z(C_{2z}k) = v_z(k) \end{cases} \quad (1)$$

$$\begin{cases} \Omega_x(C_{2z}k) = -\Omega_x(k) \\ \Omega_y(C_{2z}k) = -\Omega_y(k) \\ \Omega_z(C_{2z}k) = \Omega_z(k) \end{cases} \quad (2)$$

In the case where the laser's propagating direction is along the z-axis and the current is measured along the x or y-axis, the preserved 2-fold rotational symmetry $C_{2z}$ leads to a zero net current, since

$$\beta_{xz}(C_{2z}k) = [-v_x(k)]\Omega_z(k) = -\beta_{xz}(k) \quad (3)$$

and

$$\beta_{yz}(C_{2z}k) = [-v_y(k)]\Omega_z(k) = -\beta_{yz}(k). \quad (4)$$

- $C_{3z}$ rotational symmetry:

Under a 3-fold rotational symmetry $C_{3z}$, a generic point $\mathbf{k}$ is related to $\mathbf{k}' = C_{3z}\mathbf{k} = (-\frac{1}{2}k_x - \frac{\sqrt{3}}{2}k_y, \frac{\sqrt{3}}{2}k_x - \frac{1}{2}k_y, k_z)$ and $\mathbf{k}'' = C_{3z}^2\mathbf{k} = (-\frac{1}{2}k_x + \frac{\sqrt{3}}{2}k_y, -\frac{\sqrt{3}}{2}k_x - \frac{1}{2}k_y, k_z)$ . The Fermi velocity and Berry curvature will follow these relationships (Extended Fig. 3 (c-d)):

$$\begin{cases} v_x(C_{3z}\mathbf{k}) = -\dfrac{1}{2}v_x(\mathbf{k}) - \dfrac{\sqrt{3}}{2}v_y(\mathbf{k}) \\ v_y(C_{3z}\mathbf{k}) = \dfrac{\sqrt{3}}{2}v_x(\mathbf{k}) - \dfrac{1}{2}v_y(\mathbf{k}) \\ v_z(C_{3z}\mathbf{k}) = v_z(\mathbf{k}) \end{cases} \quad (5)$$

$$\begin{cases} \Omega_x(C_{3z}\mathbf{k}) = -\dfrac{1}{2}\Omega_x(\mathbf{k}) - \dfrac{\sqrt{3}}{2}\Omega_y(\mathbf{k}) \\ \Omega_y(C_{3z}\mathbf{k}) = \dfrac{\sqrt{3}}{2}\Omega_x(\mathbf{k}) - \dfrac{1}{2}\Omega_y(\mathbf{k}) \\ \Omega_z(C_{3z}\mathbf{k}) = \Omega_z(\mathbf{k}) \end{cases} \quad (6)$$

and

$$\begin{cases} v_x(C_{3z}^2\mathbf{k}) = -\dfrac{1}{2}v_x(\mathbf{k}) + \dfrac{\sqrt{3}}{2}v_y(\mathbf{k}) \\ v_y(C_{3z}^2\mathbf{k}) = \dfrac{\sqrt{3}}{2}v_x(\mathbf{k}) - \dfrac{1}{2}v_y(\mathbf{k}) \\ v_z(C_{3z}^2\mathbf{k}) = v_z(\mathbf{k}) \end{cases} \quad (7)$$

$$\begin{cases} \Omega_x(C_{3z}^2\mathbf{k}) = -\dfrac{1}{2}\Omega_x(\mathbf{k}) + \dfrac{\sqrt{3}}{2}\Omega_y(\mathbf{k}) \\ \Omega_y(C_{3z}^2\mathbf{k}) = \dfrac{\sqrt{3}}{2}\Omega_x(\mathbf{k}) - \dfrac{1}{2}\Omega_y(C_{3z}^2\mathbf{k}) \\ \Omega_z(C_{3z}^2\mathbf{k}) = \Omega_z(\mathbf{k}) \end{cases} \quad (8)$$

Considering that the laser propagates along z-axis and the current is measured along the x- or y-axis, the preservation of the 3-fold rotational symmetry results in a null net current because

$$\begin{cases} \beta_{xz}(C_{3z}\mathbf{k}) = \left[-\dfrac{1}{2}v_x(\mathbf{k}) - \dfrac{\sqrt{3}}{2}v_y(\mathbf{k})\right]\Omega_z(\mathbf{k}) = -\dfrac{1}{2}\beta_{xz}(\mathbf{k}) - \dfrac{\sqrt{3}}{2}\beta_{yz}(\mathbf{k}) \\ \beta_{xz}(C_{3z}^2\mathbf{k}) = \left[-\dfrac{1}{2}v_x(\mathbf{k}) + \dfrac{\sqrt{3}}{2}v_y(\mathbf{k})\right]\Omega_z(\mathbf{k}) = -\dfrac{1}{2}\beta_{xz}(\mathbf{k}) + \dfrac{\sqrt{3}}{2}\beta_{yz}(\mathbf{k}) \end{cases} \quad (9)$$

and

$$\begin{cases} \beta_{yz}(C_{3z}\mathbf{k}) = \left[\dfrac{\sqrt{3}}{2}v_x(\mathbf{k}) - \dfrac{1}{2}v_y(\mathbf{k})\right]\Omega_z(\mathbf{k}) = \dfrac{\sqrt{3}}{2}\beta_{xz}(\mathbf{k}) - \dfrac{1}{2}\beta_{yz}(\mathbf{k}) \\ \beta_{yz}(C_{3z}^2\mathbf{k}) = \left[-\dfrac{\sqrt{3}}{2}v_x(\mathbf{k}) - \dfrac{1}{2}v_y(\mathbf{k})\right]\Omega_z(\mathbf{k}) = -\dfrac{\sqrt{3}}{2}\beta_{xz}(\mathbf{k}) - \dfrac{1}{2}\beta_{yz}(\mathbf{k}). \end{cases} \quad (10)$$

## VI. Discussion on the transition temperature and domain

We note that, from the data, it appears that the CPGE onset occurs at a slightly higher temperature than $T_{CDW}$. While $T_{CDW}$ in our devices is approximately 82 K (Extended Fig. 2), slightly higher than that of the bulk crystal, the CPGE onset temperature remains noticeably higher. Although the exact reason for the difference between the CPGE onset and $T_{CDW}$ remains unclear, their close proximity suggests a possible connection between the CDW and the emergence of the CPGE. Additionally, if the first-order phase transition at $T_{CDW}$ was fluctuation driven, the observed CPGE may detect the fluctuations of the CDW order parameter, potentially explaining its onset above $T_{CDW}$. Future studies, e.g., noise spectroscopy, would be required to elucidate whether such fluctuations onset above $T_{CDW}$. On the other hand, we cannot rule out the possibility of an additional phase transition appearing at a higher temperature and further leading to a chiral charge order. Such intermediate phase has been observed in 2H-TaSe$_2$, occurring prior to the CDW onset[46]. However, establishing the existence of such a phase would require thermodynamic evidence, which is currently not available. Existing thermodynamic data do not show a clear phase transition corresponding to an intermediate state[47]. Therefore, it is more likely that the CPGE onset is linked to the CDW onset and associated fluctuations rather than to an intermediate phase whose existence remains to be confirmed through thermodynamic probes.

Based on the observation of CPGE using a 10 μm wavelength laser beam and the independence of CPGE on the induction process, we suggest that the domains within the laser-illuminated region are likely monochiral. The chiral orientation may be determined by the crystal structure, device architecture, or other factors yet to be identified, which remain as open questions. We note that in a prior study[18], an unconventional specular optical rotation angle ($\theta_C$) was observed in CsV$_3$Sb$_5$ via MOKE measurements below $T_{CDW}$. The $\theta_C$ was found to be uniformly distributed across the bulk sample, demonstrating single-domain behavior over areas exceeding $10^4 \ \mu m^2$, and remained unchanged after thermal cycling. This uniformity agrees with our findings, suggesting that the growth conditions for AV$_3$Sb$_5$ may favor specific near-degenerate chiral ground states, potentially due to A-site vacancies[39].

## VII. Magnetic field dependence of the CPGE at low temperature.

While our current experimental setup does not directly support a tunable magnetic field, we conduct a field-dependence analysis of the CPGE effect deep within the CDW state at two fixed fields. For this purpose, we placed the device on a permanent magnet, exposing it to an out-of-plane magnetic field of approximately 0.2 T (measured at room temperature). By flipping the magnet and remounting the sample, we reversed the field direction to study its influence.

Our polarization-dependence data, presented in Extended Fig. 5, demonstrates that both the magnitude and directionality of the longitudinal CPGE are modulated by $B_z$. Specifically, under a positive $B_z$, the $I_z$ generated by left-circular-polarized light (QWP angle = 45°) is noticeably weaker than that generated by right-circular-polarized light (QWP angle = 135°). Conversely, under a negative $B_z$, the $I_z$ generated by left-circular-polarized light (QWP angle = 45°) is stronger than that generated by right-circular-polarized light (QWP angle = 135°), indicating a CPGE with the opposite sign with respect to that under a positive $B_z$. It is worth noting that, at negative $B_z$, the CPGE magnitude is reduced.

While a more comprehensive field-dependent study is needed to fully unravel the role of time-reversal symmetry in the CPGE effect, our findings strongly suggest that the magnetic-field sensitivity of the CPGE and the associated time-reversal symmetry breaking are intrinsic to the CDW state, consistent with previous reports[13,14]. These results provide evidence of the interplay between magnetic fields and electronic ordering in $KV_3Sb_5$.

## VIII.     Laser power dependence of the CPGE at low temperature

We measured the effect of laser power on CPGE at the base temperature. Specifically, we varied the incident power and determined the half-difference between the photoresponses under left- and right-circularly polarized light. The results, shown in Extended Fig. 6, reveal a linear power dependence of the longitudinal CPGE. This linear relationship is expected for both second-order nonlinear effects and third-order effects facilitated by a DC electric field[31]. While the linear power dependence alone cannot definitively exclude the third-order effect, the magnetic-field-dependence of the CPGE, which is closely associated with the intrinsic time-reversal symmetry breaking characteristic of the unconventional CDW phase, supports the second-order effect (intrinsic CPGE) rather than a third-order effect driven by the out-of-plane electric-field in the graphite-$KV_3Sb_5$ interface.

## Extended Figures

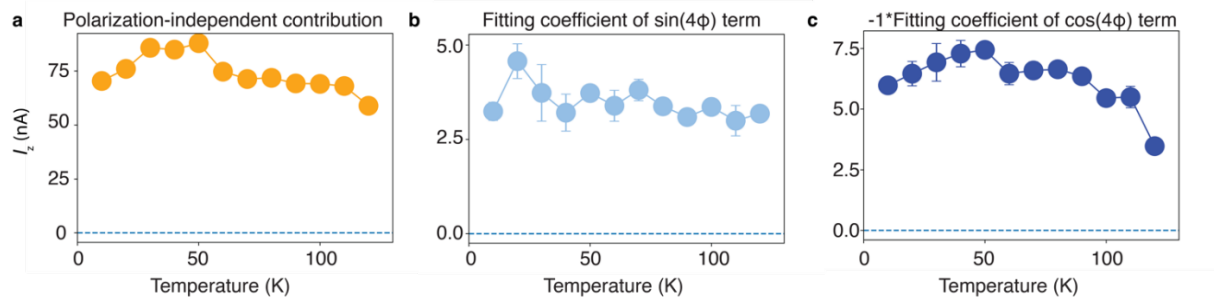

**Extended Fig. 1: Temperature dependence of two LPGE components and the polarization-independent photocurrent. a**, Photo-thermoelectric current as a function of the temperature, exhibiting no noticeable change across the CDW transition temperature. **b**, **c** Temperature dependence of two distinct LPGE contributions: component B (fitting coefficient of $\sin 4\phi$ term) and C (fitting coefficient of the $\cos 4\phi$ term), respectively. Notably, component B's value stays relatively stable throughout the entire temperature range examined, whereas component C exhibits a temperature response akin to that of the polarization-independent photocurrent. This observation suggests that components B and C have different origins, with component C aligning more closely with the behavior of the photo-thermoelectric current.

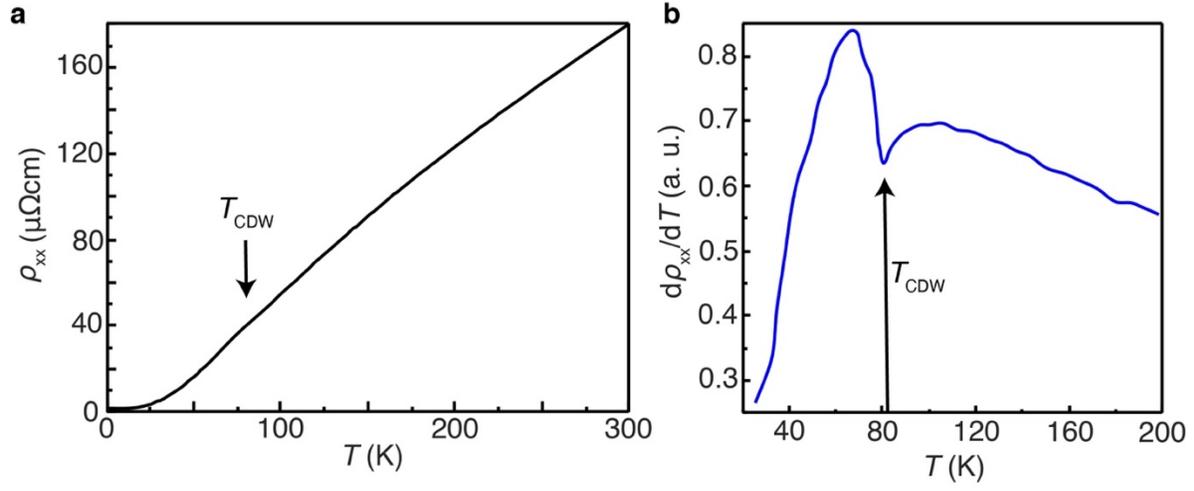

**Extended Fig. 2: Transport anomaly associated with the CDW transition in the exfoliated KV$_3$Sb$_5$ flake used for photocurrent measurements. a**, Resistivity ($\rho_{xx}$) as a function of temperature ($T$) measured in the exfoliated KV$_3$Sb$_5$ flake, revealing a phase transition towards the CDW state. **b**, Derivative of the resistivity (d$\rho_{xx}$/d$T$) with respect to temperature (from the trace in panel **a)**, highlighting a pronounced transport anomaly due to the CDW transition near $T \approx 82$ K.

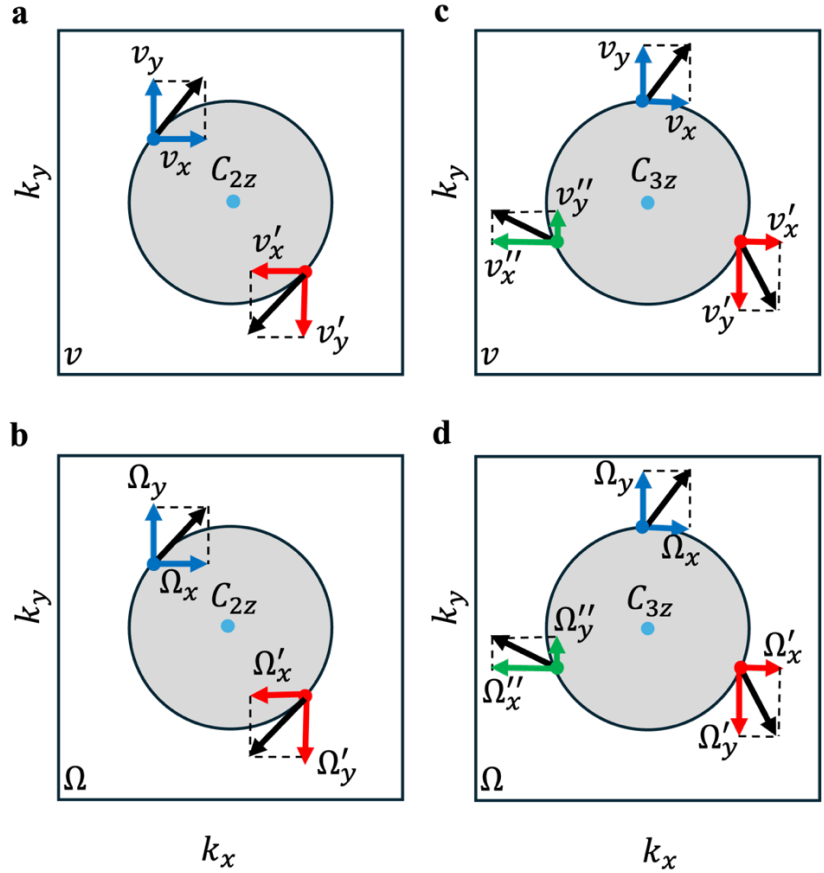

**Extended Fig. 3: Illustration of the effects of a, b. 2-fold and c, d. 3-fold rotational symmetries on velocity and Berry curvature.**

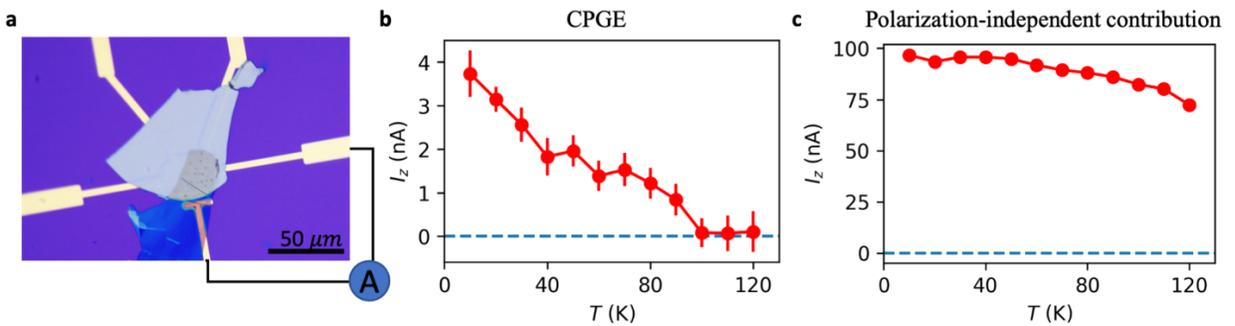

**Extended Fig. 4: Longitudinal photocurrent measured in a second device as a function of the temperature.** *a*, Optical microscopy image of the device used for collecting the out-of-plane photocurrent. *b*, Temperature dependence of the circular photogalvanic effect (CPGE), showing that the CPGE nearly vanishes above $T = 100$ K, which is consistent with the results from the first device (Fig. 2). *c*, Temperature dependence of the polarization-independent thermal component shown for comparison. While the CPGE onset temperature is reproducible among samples, subtle features below the onset appears to be sample dependent.

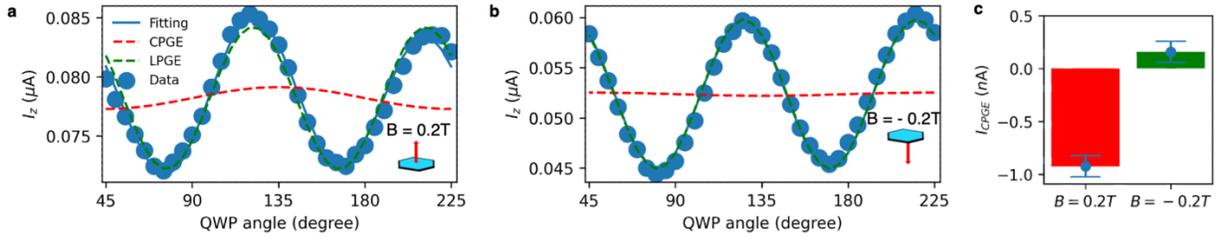

**Extended Fig. 5: CPGE in the exfoliated $KV_3Sb_5$ sample as a function of the magnetic field. a**, Polarization dependence of $I_z$ under a 0.2 T out-of-plane magnetic field applied using a permanent magnet positioned beneath the device. **b**, Polarization dependence of $I_z$ measured under the reversed magnetic field orientation, achieved by flipping the magnet. **c**, The extracted CPGE coefficients from **a** and **b**. Both measurements were performed at a temperature of 10 K.

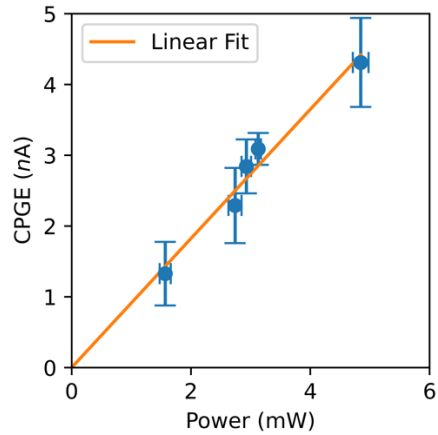

**Extended Fig. 6: Laser power dependence of the longitudinal CPGE measured on $KV_3Sb_5$ at $T = 10$ K.** Data points exhibit a linear dependence on laser power, as indicated by the orange line representing the best fit.

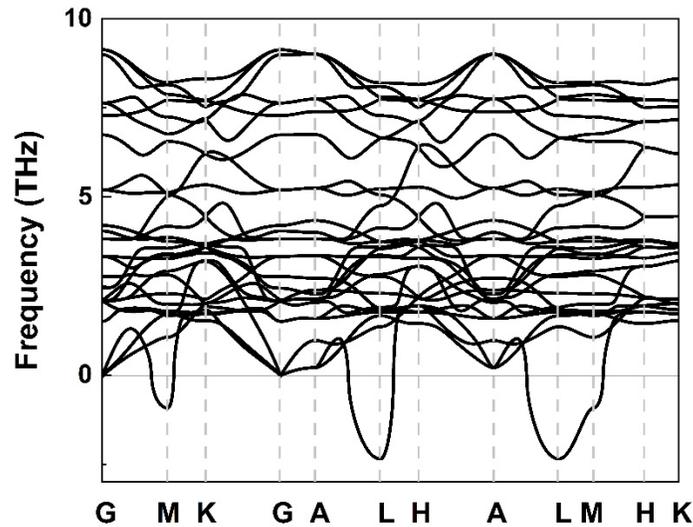

**Extended Fig. 7: Phonon spectrum of pristine KV$_3$Sb$_5$.**

**Acknowledgement:**

We acknowledge illuminating discussions with Titus Neupert. M.Z.H. group acknowledges primary support from the US Department of Energy, Office of Science, National Quantum Information Science Research Centers, Quantum Science Center (at ORNL) and Princeton University; STM Instrumentation support from the Gordon and Betty Moore Foundation (GBMF9461) and the theory work; and support from the US DOE under the Basic Energy Sciences programme (grant number DOE/BES DE-FG-02-05ER46200) for the theory and sample characterization work including ARPES. Work at Nanyang Technological University was supported by the National Research Foundation, Singapore, under its Fellowship Award (NRF-NRFF13-2021-0010), the Agency for Science, Technology and Research (A*STAR) under its Manufacturing, Trade and Connectivity (MTC) Individual Research Grant (IRG) (Grant No.: M23M6c0100), and the Nanyang Assistant Professorship grant (NTU-SUG). The sample growth was supported by the National Key Research and Development Program of China (grant nos 2020YFA0308800 and 2022YFA1403400), the National Science Foundation of China (grant no 92065109), and the Beijing Natural Science Foundation (grant nos Z210006 and Z190006). Z.W. thanks the Analysis and Testing Center at BIT for assistance in facility support. L.B. is supported by the US-DoE, Basic Energy Sciences program through award DE-SC0002613. The National High Magnetic Field Laboratory is supported by the US-NSF Divisions of Materials Research and Chemistry through DMR-2128556 and the State of Florida.